\documentclass[letterpaper]{mn2e} \usepackage{epsfig,psfig}

\title[Spatial distribution of Galactic HII regions]{Spatial distribution of Galactic HII regions}
\author[R. Paladini, R.D. Davies, G. De Zotti]
{R. Paladini$^{1}$\thanks{E-mail: paladini@sissa.it}, R. D.
Davies$^{2}$\thanks{E-mail:rdd@jb.man.ac.uk}, G.
DeZotti$^{3}$\thanks{E-mail: dezotti@pd.astro.it}\\ $^{1}$SISSA,
International School for Advanced Studies, via Beirut 2-4, I-34014
Trieste, Italy\\ $^{2}$University of Manchester, Jodrell Bank
Observatory, Macclesfield - Cheshire SK11 9DL, UK\\
$^{3}$INAF--Osservatorio Astronomico di Padova, Vicolo
dell'Osservatorio 5, I-35122 Padova, Italy}
\begin{document}

\pagerange{\pageref{firstpage}--\pageref{lastpage}} \pubyear{2003}

\maketitle

\label{firstpage}

\def\lsim{\,\lower2truept\hbox{${<\atop\hbox{\raise4truept\hbox{$\sim$}}}$}\,}
\def\gsim{\,\lower2truept\hbox{${> \atop\hbox{\raise4truept\hbox{$\sim$}}}$}\,}

\begin{abstract}
We present a new, detailed, analysis of the spatial distribution
of Galactic HII regions, exploiting a far richer database than
used in previous analyses. Galactocentric distances have been
derived for 550 objects. Distances from the Sun could be
unambiguously derived from velocity data for 117 of them, lying
either outside the solar circle (84) or on a line-of-sight
tangential to their orbit (33). For 177 further sources, distance
estimates are made possible by the use of auxiliary data. A highly
significant correlation between luminosity and linear diameter was
found and the corresponding least-square linear relationship in
the log-log plane was used to resolve the distance ambiguity for
an additional 256 sources. Within the solar circle the thickness
of the distribution of HII regions around the Galactic plane was
found to be comparable to that of OB stars (Bronfman et al. 2000).
At larger galactocentric radii the shape of the distribution
reflects that of the warp, and its thickness along the $z$ axis
increases with increasing distance from the Galactic centre. We
also confirm, for a much larger sample, the previously reported
positive gradient of electron temperature with galactocentric
distance.

\end{abstract}

\begin{keywords}
HII regions -- Galaxy: structure -- radio continuum: ISM
\end{keywords}

\section{Introduction}

As it is well known, HII regions are among the most reliable
tracers of the Galactic spiral structure (Vall\'ee 1995). The
reconstruction of their spatial distribution is also important for
understanding the distribution of free electrons (Taylor $\&$
Cordes 1993; Cordes $\&$ Lazio 2002, 2003). Recovering such a distribution,
however, is complicated by the distance degeneracy problem. While
the galactocentric distances can be derived from the Galactic
rotation curve if velocity data are available, there are in
general two solutions (``near'' and ``far'') for the distance from
the Sun of regions within the solar circle.

In this paper we exploit the rich information content of the
extensive radio catalog of Galactic HII regions published by
Paladini et al. (2003, hereafter Paper I) to address this problem.
The paper is organized as follows. In Sect.~2 we briefly describe
the catalog and analyze its completeness. In Sect.~3 we describe
the derivation of galactocentric distances and investigate the
radial gradient of the electron temperature, discovered in
previously published works. In Sect.~4 we discuss distances from
the Sun and me\-thods to overcome the degeneracy; the thickness of
the HII layer is also estimated. The main conclusions are
summarized in Sect.~5.

\section[]{The catalog}

\subsection[]{Description}

The extensive radio catalog of Galactic HII regions, presented in
Paper I, has been produced by exploiting the data contained in 24
previously published lists (see Table~1 in Paper I for details and
references). The final compilation - the so-called Master Catalog
- includes 1442 classical HII regions (i.e., bright, compact
objects) for which original flux densities and diameters as well
as the available information on line velocities, line widths and
line temperatures and the errors on these quantities are listed.
The catalog does not include ultra-compact (UCHII) and extremely
extended (EHE) sources.

A Synthetic Catalog of flux densities and diameters at 2.7 GHz
(with the corresponding errors) for each of the 1442 sources of
the Master Catalog has also been produced. When not directly
available, the flux density - and the corresponding error - at 2.7
GHz has been computed by extrapolating/interpolating the published
observational data at other frequencies. For sources lacking a
measurement of the angular diameter (14$\%$ of the total), an
indicative diameter has been estimated, based on the observed
angular diameter - flux density correlation (see Paper I for more
details).

\subsection[]{Completeness of the catalog}

As shown by Fig.~1, the bright tail (300 Jy $\gsim \hskip 0.1truecm S_{2.7{\rm GHz}}
\hskip 0.1truecm \gsim$ 70 Jy)
of the integral counts, $N(>S)$, of HII regions listed in Paper I exhibits the
Euclidean slope [$N(>S)\propto S^{-3/2}$] as expected for the
nearest sources (within a distance not exceeding the thickness of
the HII layer). At fainter fluxes the counts flatten to
$N(>S)\propto S^{-1}$, consistent with the 2-D (disk-like)
distribution of more distant sources. Below $\simeq 7\,$Jy the
counts become still flatter, suggesting the onset of a substantial
incompleteness.

The completeness limit is to a large extent determined by source
confusion. The surface density of catalogued sources reaches
values of one source per few beams in the region $\pm 60^\circ$
around $l=0^\circ$, $b \sim 0^\circ$, for a typical $10'$ beam. As
extensively discussed in the literature (e.g., Scheuer 1957;
Condon 1974; Hogg 2001) confusion becomes important at flux levels
at which there are more than $1/30$ sources per beam.
Correspondingly, it is very difficult to resolve into individual
sources the structure seen right in the Galactic plane,
particularly towards the Galactic center. Thus, the majority of
weak HII regions are actually found in less crowded areas, such as
the anti-center region or regions at $|b|>1^\circ$.

\begin{figure}
\mbox{} \epsfig{figure=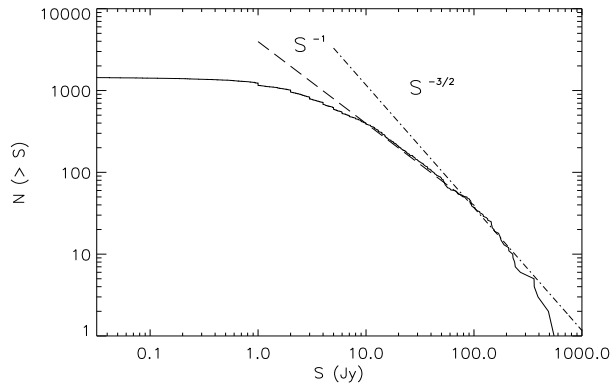,
height=5cm, width=8cm, angle=0} \caption{Cumulative counts N($>
S$) of HII regions in the
catalog by Paladini et al. (2003).}

\end{figure}

\begin{figure}
\epsfig{figure=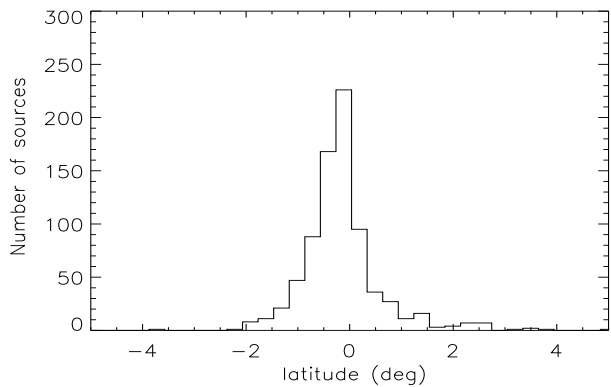,
height=5cm, width=8cm,angle=0}
 \caption{Galactic latitude distribution of $\simeq$ 800 sources with detection of recombination
lines.}
\end{figure}

\section[]{Galactocentric distances of HII regions}

\subsection{Distance estimates}

Paper I contains radio recombination line velocities for $\simeq
800$ of the catalogued HII regions. The references for the
velocity data and the lines used are listed in Table~1. This is,
by far, the largest sample used to derive distances. The present
analysis also improves on the earlier ones because of the adoption
of an updated Galaxy rotation curve. Previous studies (Caswell \&
Haynes 1987; Downes et al. 1980) were based on $\simeq 300$
sources and used Schmidt's (1965) rotation curve or slightly
modified versions of it.

\begin{table}
\caption {References for velocity data.}
\begin{tabular}{lcc}
\hline
\hline
{\hskip 1.1truecm {\it Reference }}   &    {\it Line}\\
\hline
Caswell $\&$ Haynes 1987 & H109$\alpha$/H110$\alpha$\\
Downes et al. 1980 & H110$\alpha$\\
Lockman 1989 & H85$\alpha$/H87$\alpha$/H88$\alpha$\\
Reifenstein et al. 1970 &  H109$\alpha$\\
Wilson et al. 1970 &  H109$\alpha$\\
Wink et al. 1982 & H76$\alpha$/H90$\alpha$\\
Wink et al. 1983 & H76$\alpha$\\
\hline
\hline
\end{tabular}
\end{table}

\begin{figure}
\mbox{}
\epsfig{figure=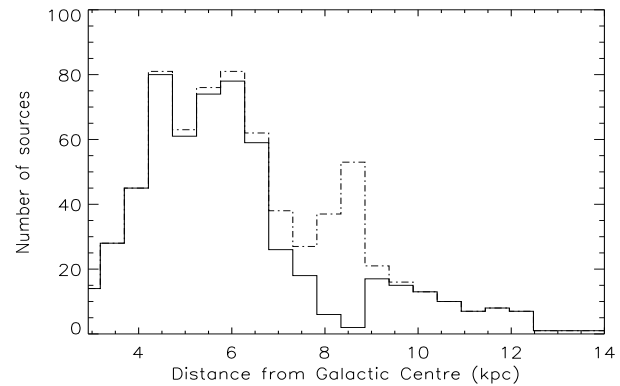,
height=5cm, width=8cm, angle=0}
 \caption {Distribution of
galactocentric distances of all $\simeq$ 800 sources with velocity data (dashed 
line). The solid line shows the distribution of 575 sources with
radial velocities  $|V_{r}| > 10 \hbox{km}\,\hbox{s}^{-1}$.}

\end{figure}

The rotation velocity $\Theta$ around the Galactic centre of an
object at galactocentric distance $R$,  Galactic longitude $l$,
and with radial velocity $V_{r}$ in the local standard of rest, is
given by:

\begin{equation}
\Theta = (R/R_{0})(\Theta_{0}+V_{r}/\sin l)
\end{equation}

\noindent where $\Theta_{0}$ and $R_{0}$ denote, respectively, the
rotation velocity and the galactocentric distance of the Sun. We
adopt the IAU-recommended values of $R_{0} = 8.5\,$kpc and
$\Theta_{0} = 220\,\hbox{km}\,\hbox{s}^{-1}$ (Kerr $\&$
Lynden-Bell 1986). From Eq.~(1) we can derive $R$ from radial
velocity measurements, if the rotation curve $\Theta(R)$ is known.
We have used the linear expression derived by Fich, Blitz and
Stark (1989) (hereafter FBS89), holding for galactocentric
distances between 3 and 17 kpc:

\begin{equation}
\Theta = (221.64 - 0.44 R) \,\hbox{km}\,\hbox{s}^{-1}
\end{equation}

\noindent where $R$ is in kpc.

The outer part of the FBS89 rotation curve was constructed  using 
HII regions spectrophotometric distances and velocities. Our sample 
has some sources in common with theirs, namely those with CO velocity 
measurements. To avoid 
circularity we have excluded such sources from further analysis. Thus 
the velocity measurements we have used are fully independent 
of those used by FBS89. 
As discussed by Blitz (1979) and FBS89, the use of HII regions to derive 
a rotation curve yields relatively small absolute errors compared with 
methods relying  on other classes of objects such as star clusters, 
planetary nebulae, carbon stars, and diffuse atomic hydrogen. On the other 
hand, the good agreement of the FBS89 rotation curve with determinations 
using other methods indicates that systematic errors specific to 
HII regions cannot be large except, perhaps, in specific regions 
where systematic, non-circular, velocity components (e.g. streaming motions)  
are present. One such region is the Perseus arm where a mean streaming 
velocity $\simeq$ 12 km sec$^{-1}$ has been found by Brand \& Blitz (1993). 
As a consequence, the distances of 36 sources in that region 
may be affected by a substantial systematic error.

The mean measurement error on $V_r$ is $\simeq
1\,\hbox{km}\,\hbox{s}^{-1}$, although with a large scatter (the
minimum error is $0.02\,\hbox{km}\,\hbox{s}^{-1}$; the maximum
$10.7\,\hbox{km}\,\hbox{s}^{-1}$). The corresponding error on the
derived values of $R$ are typically of order 1\%. On the other
hand, random
motions  can add a significant uncertainty to the computed
values of R. The average local peculiar velocities reported by
Stark \& Brand (1989) and Clemens (1985) are $\simeq
5\,\hbox{km}\,\hbox{s}^{-1}$. To curtail this effect, we
consider only objects with $|V_r| \ge
10\,\hbox{km}\,\hbox{s}^{-1}$. As a consequence, we are left with
575 sources.

The distribution of the derived galactocentric distances is shown
in Fig.~3, where the peaks corresponding to the spiral arms at $R$
= 4, 6 and 8 kpc are visible. The latter peak, due to sources on
or near the solar circle, however disappears if we consider only
sources with $|V_r| \ge 10\,\hbox{km}\,\hbox{s}^{-1}$. 

\subsection{Electron temperature vs Galactocentric distance}

We are now in a position to investigate the dependence of HII region
electron temperatures upon Galactocentric distance. 
For 404 of the 575 sources with more reliable estimates of $R$, electron
temperatures, $T_e$, are available (Caswell $\&$ Haynes 1987;
Downes et al. 1980; Reifenstein et al. 1970; Wilson et al. 1970;
Wink et al. 1982, 1983) and listed in Paper I. Non-LTE effects
should be small over the entire range of frequencies 5 to 22 GHz
(Shaver et al.1983; Wink et al.~1983), so that corrections to the
computed $T_{e}$ values are not ne\-ces\-sa\-ry. When two or more
values of $T_e$ are available for the same source, their weighted
average has been taken. The data shown in Fig.~4 confirm the
previously reported correlation between $T_e$ and $R$, based on
much smaller samples (Churchwell et al. 1978; Downes et al. 1980;
Shaver et al. 1983). The least-square linear relationship is:
\begin{equation}
T_{e} = (4166 \pm 124) + (314 \pm 20) R \hspace*{0.2truecm}
\hbox{K}\
\end{equation}
\noindent close to the previous result by Shaver et al. (1983) who
found $T_{e} = (3150 \pm 110) + (433 \pm 40) R$ \hspace*{0.2truecm} K.

\begin{figure}
{\epsfig{figure=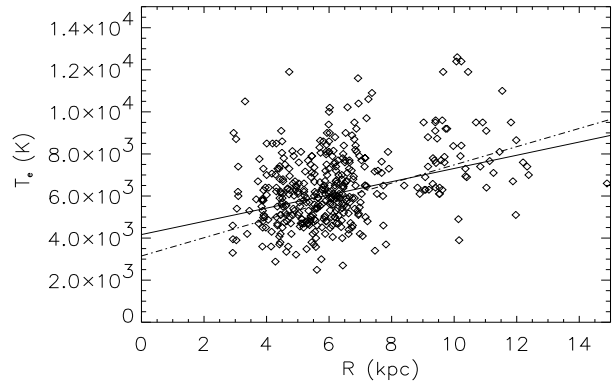,
height=5cm, width=8cm, angle=0}} \caption[]{Electron temperatures
versus galactocentric distance. The solid line shows our
least-squares linear relationship, the dot-dashed line that found
by Shaver et al. (1983).}
\end{figure}

\section[]{Distances from the Sun}

The galactocentric radius $R$ is related to the distance $D$ from
the Sun by the equation:
\begin{equation}
R = (R_{0}^2+D^2-2 D R_{0} \cos l)^{\frac{1}{2}} \ ,
\end{equation}
\noindent which has obviously, in general, two real solutions for
$D$ if $R \le R_0$:
\begin{equation}
D_{far/near} = R_{0} (\cos l \pm ({\cos l}^2 - (1 -
(R/R_{0})^{2}))^{\frac{1}{2}})
\end{equation}
\noindent corresponding to the two intersections of the
line-of-sight with the orbit of radius $R$. The solution is unique
only for sources lying outside the solar circle or whose
line-of-sight is tangent to their orbit.

To break the degeneracy we need auxiliary data. Two kinds of data
have been used: absorption lines and optical counterparts. Since
only neutral (HI) or molecular (H$_{2}$CO and OH) gas in front of an
HII region can absorb its continuum emission, if the region is at
the ``near'' distance the gas absorption line velocity is
lower than its recombination line velocity. On the other hand, if
the HII region is at the ``far'' distance it is possible to
detect absorption at a velocity higher than that of its
recombination line. At the same time, if an optical counterpart is
detected, then it is very likely that the HII region is at the
``near'' distance since the heavy dust obscuration at optical
wavelengths makes detections of far sources quite difficult.

Table~2 lists the sources for which the ambiguity can be resolved
using auxiliary data. HI data are mainly taken from Kuchar $\&$
Bania (1994) and Caswell et al. (1975). Additional data are taken
from Kerr $\&$ Knapp (1970) and Goss $\&$ Radhakrishnan (1969). H$_{2}$CO
data are from Wilson (1980). Optical identifications are from the
catalogs by Mars\'alkov\'a (1974), Blitz et al. (1982), and
Brand $\&$ Blitz (1993). Complementary data on individual sources
are from Miller (1968) and Shaver et al. (1981).

\begin{table*}
\caption[InpVal] {Galactic HII regions for which the distance
ambiguity was resolved using auxiliary data. The $(o)$ denotes HII
regions with an optical counterpart, for which the ``near''
solution was chosen; $(a1)$ and $(a2)$ denote sources for which,
respectively, HI or H$_{2}$CO absorption data are available.}

\begin{tabular}{cccc| cccc | cccc}
\hline
\hline
{\it l}  &  {\it b} & {\it Near or Far} &  & {\it l}  &  {\it b} & {\it Near or Far} &  &{\it l}  &  {\it b} & {\it Near or Far} &  \\
&          & {\it  kinematic distance}  &  &            &          & {\it kinematic distance}   & &          &          & {\it
kinematic distance} & \\
\hline
  6.4   &  -0.5   &  near \hskip 0.3truecm   & (o) & 25.3   &   0.3   &  near \hskip 0.3truecm   & (o) & 44.3   &   0.1   &  near \hskip 0.3truecm   & (a1)\\
  6.5   &   0.1   &  near \hskip 0.3truecm   & (a1) & 25.4   &  -0.3   &  far  \hskip 0.3truecm   & (a2) & 45.1   &   0.1   &  far  \hskip 0.3truecm   & (a1)\\
  6.6   &  -0.3   &  near \hskip 0.3truecm   & (o) & 25.4   &  -0.2   &  far  \hskip 0.3truecm   & (a2) & 45.5   &   0.1   &  far  \hskip 0.3truecm   & (a1)\\
  6.6   &  -0.1   &  near \hskip 0.3truecm   & (a1) & 25.7   &   0.0   &  far  \hskip 0.3truecm    & (a2) & 46.5   &  -0.2   &  near \hskip 0.3truecm   & (a1)\\
  6.7   &  -0.2   &  near \hskip 0.3truecm   & (a1) & 25.8   &   0.2   &  tangent \hskip 0.3truecm   & (a2) & 48.6   &   0.0   &  far  \hskip 0.3truecm   & (a1)\\
  7.0   &  -0.2   &  near \hskip 0.3truecm   & (o) & 26.1   &  -0.1   &  far  \hskip 0.3truecm   & (a2) & 48.6   &   0.2   &  far  \hskip 0.3truecm   & (a1)\\
  8.4   &  -0.3   &  near \hskip 0.3truecm   & (a2) & 27.3   &   0.1   &  far  \hskip 0.3truecm   & (a2) & 49.4   &  -0.2   &  far  \hskip 0.3truecm   & (a1)\\
  8.7   &  -0.4   &  near \hskip 0.3truecm   & (o) & 27.5   &   0.2   &  far  \hskip 0.3truecm   & (a2) & 51.1   &   0.2   &  far  \hskip 0.3truecm   & (a1)\\
 10.5   &   0.0   &  near \hskip 0.3truecm   & (a2) & 28.6   &   0.0   &  far  \hskip 0.3truecm   & (a2) & 52.8   &   0.3   &  far  \hskip 0.3truecm   & (a1)\\
 11.7   &  -1.7   &  near \hskip 0.3truecm   & (o) & 28.7   &   0.0   &  far  \hskip 0.3truecm   & (a2) & 53.6   &   0.0   &  near \hskip 0.3truecm   & (o)\\
 12.4   &  -1.1   &  near \hskip 0.3truecm   & (o) & 28.8   &   0.2   &  tangent \hskip 0.3truecm   & (a2) & 53.6   &   0.2   &  far  \hskip 0.3truecm   & (a1)\\
 13.4   &   0.1   &  far  \hskip 0.3truecm   & (a2) & 29.0   &  -0.6   &  near \hskip 0.3truecm   & (o) & 60.9   &  -0.1   &  near \hskip 0.3truecm   & (o)\\
 14.0   &  -0.1   &  near \hskip 0.3truecm   & (a2) & 29.1   &  -0.7   &  near \hskip 0.3truecm   & (o) & 62.9   &   0.1   &  near \hskip 0.3truecm   & (o)\\
 14.2   &  -0.2   &  near \hskip 0.3truecm   & (o) & 29.9   &  -0.0   &  far  \hskip 0.3truecm   & (a2) & 63.2   &   0.4   &  near \hskip 0.3truecm   & (o)\\
 14.2   &  -0.1   &  near \hskip 0.3truecm   & (o) & 30.2   &  -0.1   &  far  \hskip 0.3truecm   & (a2) & 63.2   &   0.5   &  near \hskip 0.3truecm   & (o)\\
 14.4   &  -0.7   &  near \hskip 0.3truecm   & (a2) & 30.5   &   0.0   &  near \hskip 0.3truecm   & (o) & 302.8   &   1.3   &  near \hskip 0.3truecm   & (o)\\
 14.6   &   0.0   &  near \hskip 0.3truecm   & (o) & 30.5   &   0.4   &  near \hskip 0.3truecm   & (o) & 305.4   &   0.2   &  far  \hskip 0.3truecm   & (a1)\\
 14.6   &   0.1   &  near \hskip 0.3truecm   & (a2) & 30.6   &  -0.1   &  tangent \hskip 0.3truecm   & (a1) & 308.7   &   0.6   &  near \hskip 0.3truecm   & (o)\\
 15.0   &  -0.7   &  near \hskip 0.3truecm   & (o) & 30.7   &  -0.3   &  tangent \hskip 0.3truecm   & (a2) & 311.0   &   0.4   &  near \hskip 0.3truecm   & (o)\\
 15.1   &  -0.9   &  near \hskip 0.3truecm   & (o) & 30.8   &  -0.0   &  near \hskip 0.3truecm   & (a1) & 311.9   &   0.1   &  far  \hskip 0.3truecm   & (a1)\\
 15.1   &  -0.7   &  near \hskip 0.3truecm   & (o) & 31.0   &   0.0   &  far \hskip 0.3truecm   & (a1) & 311.9   &   0.2   &  far  \hskip 0.3truecm   & (a1)\\
 15.2   &  -0.8   &  near \hskip 0.3truecm   & (a2) & 31.2   &  -0.1   &  far  \hskip 0.3truecm   & (a1) & 316.8   &  -0.1   &  near  \hskip 0.3truecm   & (o)\\
 15.2   &  -0.6   &  near \hskip 0.3truecm   & (a2) & 31.3   &   0.1   &  tangent  \hskip 0.3truecm   & (a1) & 316.8   &  -0.0   &  near \hskip 0.3truecm   & (o)\\
 16.6   &  -0.3   &  near \hskip 0.3truecm   & (o) & 31.4   &  -0.3   &  near \hskip 0.3truecm   & (a2) & 317.0   &   0.3   &  far  \hskip 0.3truecm   & (1a)\\
 16.9   &   0.8   &  near \hskip 0.3truecm   & (o) & 31.4   &   0.3   &  tangent \hskip 0.3truecm   & (a2) & 320.2   &   0.8   &  near \hskip 0.3truecm   & (o)\\
 17.0   &   0.8   &  near \hskip 0.3truecm   & (o) & 31.6   &   0.1   &  near \hskip 0.3truecm   & (a1) & 321.1   &  -0.5   &  near \hskip 0.3truecm   & (o)\\
 17.0   &   0.9   &  near \hskip 0.3truecm   & (a2) & 31.8   &   1.5   &  near \hskip 0.3truecm   & (o) & 322.2   &   0.6   &  near \hskip 0.3truecm   & (a1)\\
 18.1   &  -0.3   &  near \hskip 0.3truecm   & (a2) & 32.2   &   0.1   &  far  \hskip 0.3truecm   & (a1) & 324.2   &   0.1   &  near \hskip 0.3truecm   & (o)\\
 18.2   &  -0.3   &  near \hskip 0.3truecm   & (o) & 32.8   &   0.2   &  far  \hskip 0.3truecm   & (a1) & 326.6   &   0.6   &  near \hskip 0.3truecm   & (a1)\\
 18.2   &  -0.4   &  near \hskip 0.3truecm   & (a2) & 33.1   &  -0.1   &  tangent  \hskip 0.3truecm   & (a1) & 326.7   &   0.6   &  near \hskip 0.3truecm   & (o)\\
 18.3   &  -0.4   &  near \hskip 0.3truecm   & (a2) & 33.4   &  -0.0   &  far  \hskip 0.3truecm   & (a1) & 327.3   &  -0.6   &  near \hskip 0.3truecm   & (a1)\\
 18.3   &  -0.3   &  near \hskip 0.3truecm   & (a2) & 34.3   &   0.1   &  near \hskip 0.3truecm   & (a1) & 327.3   &  -0.5   &  near \hskip 0.3truecm   & (o)\\
 18.3   &   1.9   &  near \hskip 0.3truecm   & (a2) & 34.9   &  -0.0   &  near \hskip 0.3truecm   & (a1) & 328.0   &  -0.1   &  near \hskip 0.3truecm   & (o)\\
 18.7   &   2.0   &  near \hskip 0.3truecm   & (o) & 35.1   &  -1.5   &  near \hskip 0.3truecm   & (a2) & 328.3   &   0.4   &  far  \hskip 0.3truecm   & (a1)\\
 18.9   &  -0.4   &  near \hskip 0.3truecm   & (o) & 35.2   &  -1.8   &  near \hskip 0.3truecm   & (a2) & 328.6   &  -0.5   &  near \hskip 0.3truecm   & (o)\\
 18.9   &  -0.5   &  near \hskip 0.3truecm   & (a2) & 35.3   &  -1.8   &  near \hskip 0.3truecm   & (a2) & 330.9   &  -0.4   &  near \hskip 0.3truecm   & (a1)\\
 19.0   &  -0.0   &  near \hskip 0.3truecm   & (a2) & 35.6   &  -0.5   &  near \hskip 0.3truecm   & (a1) & 331.3   &  -0.3   &  near \hskip 0.3truecm   & (a1)\\
 19.1   &  -0.3   &  near \hskip 0.3truecm   & (a2) & 35.6   &  -0.0   &  far  \hskip 0.3truecm   & (a1) & 331.5   &  -0.1   &  far  \hskip 0.3truecm   & (a1)\\
 19.6   &  -0.2   &  near \hskip 0.3truecm   & (a2) & 35.6   &   0.1   &  far  \hskip 0.3truecm   & (a1) & 332.8   &  -1.4   &  near \hskip 0.3truecm   & (o)\\
 19.6   &  -0.1   &  near \hskip 0.3truecm   & (a2) & 35.7   &  -0.0   &  near \hskip 0.3truecm   & (a2) & 332.8   &  -0.6   &  near \hskip 0.3truecm   & (o)\\
 19.7   &  -0.1   &  near \hskip 0.3truecm   & (a2) & 36.3   &  -1.7   &  near \hskip 0.3truecm   & (o) & 333.0   &  -0.4   &  far  \hskip 0.3truecm   & (a1)\\
 21.0   &   0.1   &  far  \hskip 0.3truecm   & (a2) & 36.5   &  -0.2   &  far  \hskip 0.3truecm   & (a1) & 333.1   &  -0.4   &  near \hskip 0.3truecm   & (o)\\
 21.9   &   0.0   &  near \hskip 0.3truecm   & (o)  & 37.4   &  -0.2   &  far  \hskip 0.3truecm   & (a2) & 333.3   &  -0.4   &  near \hskip 0.3truecm   & (o)\\
 22.8   &  -0.5   &  far  \hskip 0.3truecm   & (a2) & 37.4   &  -0.1   &  far  \hskip 0.3truecm   & (a1) & 333.6   &  -0.2   &  near \hskip 0.3truecm   & (a1)\\
 22.9   &  -0.3   &  far  \hskip 0.3truecm   & (a2) & 37.4   &  -0.0   &  far  \hskip 0.3truecm   & (a1) & 336.4   &  -0.2   &  near \hskip 0.3truecm   & (o)\\
 22.9   &   0.7   &  near \hskip 0.3truecm   & (o) & 37.5   &  -0.1   &  far  \hskip 0.3truecm   & (a1) & 336.5   &  -1.5   &  near \hskip 0.3truecm   & (a1)\\
 23.0   &  -0.4   &  far  \hskip 0.3truecm   & (a2) &  37.6   &  -0.1   &  far  \hskip 0.3truecm   & (a1) & 336.8   &   0.0   &  far  \hskip 0.3truecm   & (a1)\\
 23.1   &   0.6   &  near \hskip 0.3truecm   & (o) & 37.7   &  -0.1   &  far  \hskip 0.3truecm   & (a1) & 337.1   &  -0.2   &  far  \hskip 0.3truecm   & (a1)\\
 23.4   &  -0.2   &  tangent \hskip 0.3truecm   & (a2) & 37.8   &  -0.2   &  far  \hskip 0.3truecm   & (a1) & 337.9   &  -0.5   &  near \hskip 0.3truecm   & (a1)\\
 23.5   &  -0.0   &  far  \hskip 0.3truecm   & (a2) & 37.9   &  -0.4   &  far  \hskip 0.3truecm   & (a1) & 338.9   &   0.6   &  near \hskip 0.3truecm   & (a1)\\
 23.7   &   0.2   &  tangent \hskip 0.3truecm   & (a2) &  38.1   &  -0.0   &  far  \hskip 0.3truecm   & (a1) & 340.8   &  -1.0   &  near \hskip 0.3truecm   & (a1)\\
 23.9   &  -0.1   &  far  \hskip 0.3truecm   & (a2) & 39.9   &  -1.3   &  near \hskip 0.3truecm   & (o) & 345.4   &  -0.9   &  near \hskip 0.3truecm   & (a1)\\
 24.0   &   0.2   &  far  \hskip 0.3truecm   & (a2) & 40.5   &   2.5   &  near \hskip 0.3truecm   & (o) & 345.4   &   1.4   &  near \hskip 0.3truecm   & (a1)\\
 24.2   &  -0.1   &  far  \hskip 0.3truecm   & (a2) & 41.1   &  -0.2   &  far  \hskip 0.3truecm   & (a1) & 348.2   &  -1.0   &  near \hskip 0.3truecm   & (o)\\
 24.4   &   0.1   &  tangent \hskip 0.3truecm   & (a2) & 41.2   &   0.4   &  near \hskip 0.3truecm   & (a1) & 348.7   &  -1.0   &  near \hskip 0.3truecm   & (o)\\
 24.5   &   0.2   &  tangent \hskip 0.3truecm   & (a2) & 41.5   &   0.0   &  far  \hskip 0.3truecm   & (a1) & 350.5   &   1.0   &  near \hskip 0.3truecm   & (o)\\
 24.5   &   0.5   &  tangent \hskip 0.3truecm   & (a2) & 42.1   &  -0.6   &  near \hskip 0.3truecm   & (a1) & 351.6   &  -1.3   &  near \hskip 0.3truecm   & (o)\\
 24.7   &  -0.2   &  tangent \hskip 0.3truecm   & (a2) & 42.4   &  -0.3   &  far  \hskip 0.3truecm   & (a1) &\\
 24.7   &  -0.1   &  near \hskip 0.3truecm   & (o) & 42.6   &  -0.1   &  near \hskip 0.3truecm   & (a1) &\\
 24.8   &   0.1   &  tangent  \hskip 0.3truecm   & (a1) & 43.9   &  -0.8   &  far  \hskip 0.3truecm   & (a1) &\\
\hline
\hline
\end{tabular}
\end{table*}

\noindent Of our 575 sources with $|V_r| \ge
10\,\hbox{km}\,\hbox{s}^{-1}$, 117 have a unique solution for $D$
(84 lie outside the solar circle and for 33 the line of sight is
tangent to their galactocentric orbit); for 177 others we have
additional data allowing us to discriminate between the two
solutions; 281 are left with the distance degeneracy. For the
latter, we need an additional distance indicator.

\subsection{The luminosity-physical diameter correlation}

A rough proportionality between luminosity and linear diameter 
is expected based on the following argument. The luminosity 
is proportional to the product of the emission measure (which 
is proportional to the linear diameter) with the square of the 
angular size. But, for our sample, the distribution of angular sizes has a 
dispersion of only $6'$ which is small enough not to swamp 
the correlation with the emission measure.

In order to verify the correlation,
we have selected from the catalog by Paladini et al. (2003) the HII
regions with flux densities, $S_\nu$, and angular diameters,
$\theta$, measured with the same instrument as well as with an
unambiguous determination of the distance, $D$, from the Sun. We
have 57 such sources with measurements at 2.7 GHz and 190 with
measurements at 5 GHz. At both frequencies we find a highly
significant correlation between the luminosity $L_{\nu}=4\pi D^2 S_\nu$ and 
linear diameter (see Fig.~5). The Pearson
correlation coefficients are 0.54 (corresponding to a probability
$p\simeq 10^{-5}$ that the correlation is occurring by chance) at 2.7 GHz,
and 0.56 ($p\simeq 10^{-17}$) at 5 GHz.

On the other hand, we should worry about the possibility that the
correlation is an artifact arising through the dependence of
luminosity on $D^2$ while the linear diameter is proportional to
$D$. To check if this can be the case, we have computed the
partial correlation coefficient between the two quantities, i.e.
the correlation at constant $D$. For the combined sample of
$190+57$ sources (extrapolating to 5 GHz the 2.7 GHz fluxes with a
spectral index of $-0.1$, $S \propto \nu^{-0.1}$, as appropriate
for optically thin free-free emission) we find a partial
correlation coefficient of 0.37 for which $p\simeq 10^{-8}$. The
correlation is therefore clearly physically significant. This
issue will be further discussed in Sect.~5.

Assuming a linear relationship between $\log L_{\nu}$ (in erg/s)
and $\log d$ (in pc):
\begin{equation}
\log \hskip 0.1truecm L_{\nu} = a + b \times \log \hskip 0.1truecm d
\end{equation}
a least square fit yields $a=31.9$, $b=1.05$ at $2.7\,$GHz and
$a=32.4$, $b=0.86$ at $5\,$GHz. Combining the two samples, we
obtain $a=32.1$, $b=0.88$ at $5\,$GHz.

The solar distance $D$ can then be estimated as:
\begin{equation}
D = 10^{\frac{a}{2-b}}
\left(\frac{\theta}{1^{"}}\right)^{\frac{b}{2-b}}
\left(\frac{\nu}{1 GHz}\right)^{\frac{-1}{2-b}}\left(\frac{S}{1
Jy}\right)^{\frac{-1}{2-b}} \hskip 0.1truecm \hbox{kpc}
\end{equation}

Although the dispersion around the above relationship is too large
to make it a good distance indicator for individual sources, it
allows us to discriminate in a statistical sense among the ``near"
and ``far'' solutions. To test the reliability of this approach we
have applied it to the 177 sources for which the distance
degeneracy has been broken using complementary data. We find that
the fractions of ``near'' ($\simeq 65\%$) and ``far'' ($\simeq
35\%$) distances are correctly reproduced although the correct
distance is assigned to only $\simeq 60\%$ of individual sources.
\begin{figure}
\centerline{\psfig{figure=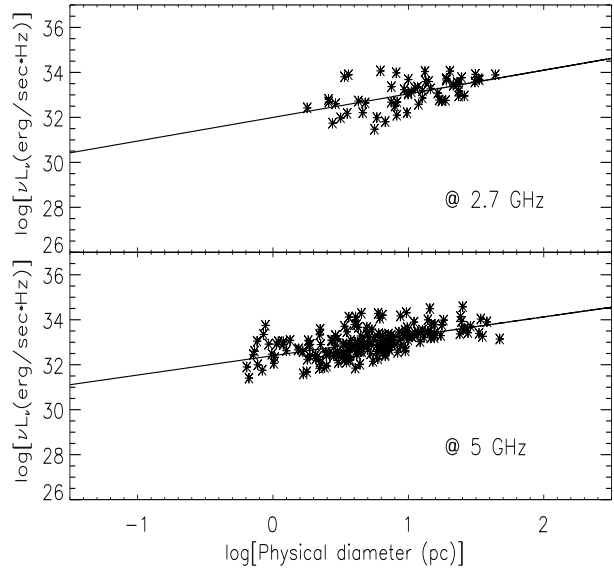,
height=7.5cm, width=8cm, angle=0}} \caption{Luminosity-physical
diameter correlation at 2.7 (57 sources) and 5 GHz (190 sources) for HII regions with
unambiguous distances.}
\end{figure}

We have homogeneous measurements of flux densities and of angular
diameters at either 2.7 or 5 GHz for 256 out of the 281 HII
regions with distance degeneracy. 
To these sources we apply Eq.~(7). We find that 100 ($\sim$ 40$\%$) 
objects are assigned to the ``near'' solution and 155 ($\sim$ 60$\%$) 
to the ``far'' solution.

\begin{table}
\caption {Fractions of near (N), far (F), and tangent (T)
solutions resulting from absorption data. The last column shows the
number of sources of the catalog whose distance ambiguity is resolved
through listed HI absorption data.}
\begin{tabular}{llcccc}
\hline \hline
                &                                   &   N ($\%$)   &     F ($\%$)  &    T ($\%$) &  N\\
\hline
                & Kuchar $\&$ Bania (1994)             &      20              &     72    &     8  &  40     \\
     HI         & Caswell et al. (1975)                &      67              &     28    &     5  &  18     \\
                & Kerr $\&$ Knapp (1970)               &      55               &    45    &     -  &   9     \\
\hline
 H$_{2}$CO      & Wilson (1980)                      &      44             &    37     &     19  &  84   \\
\hline \hline
\end{tabular}
\end{table}

\begin{figure}
{\epsfig{figure=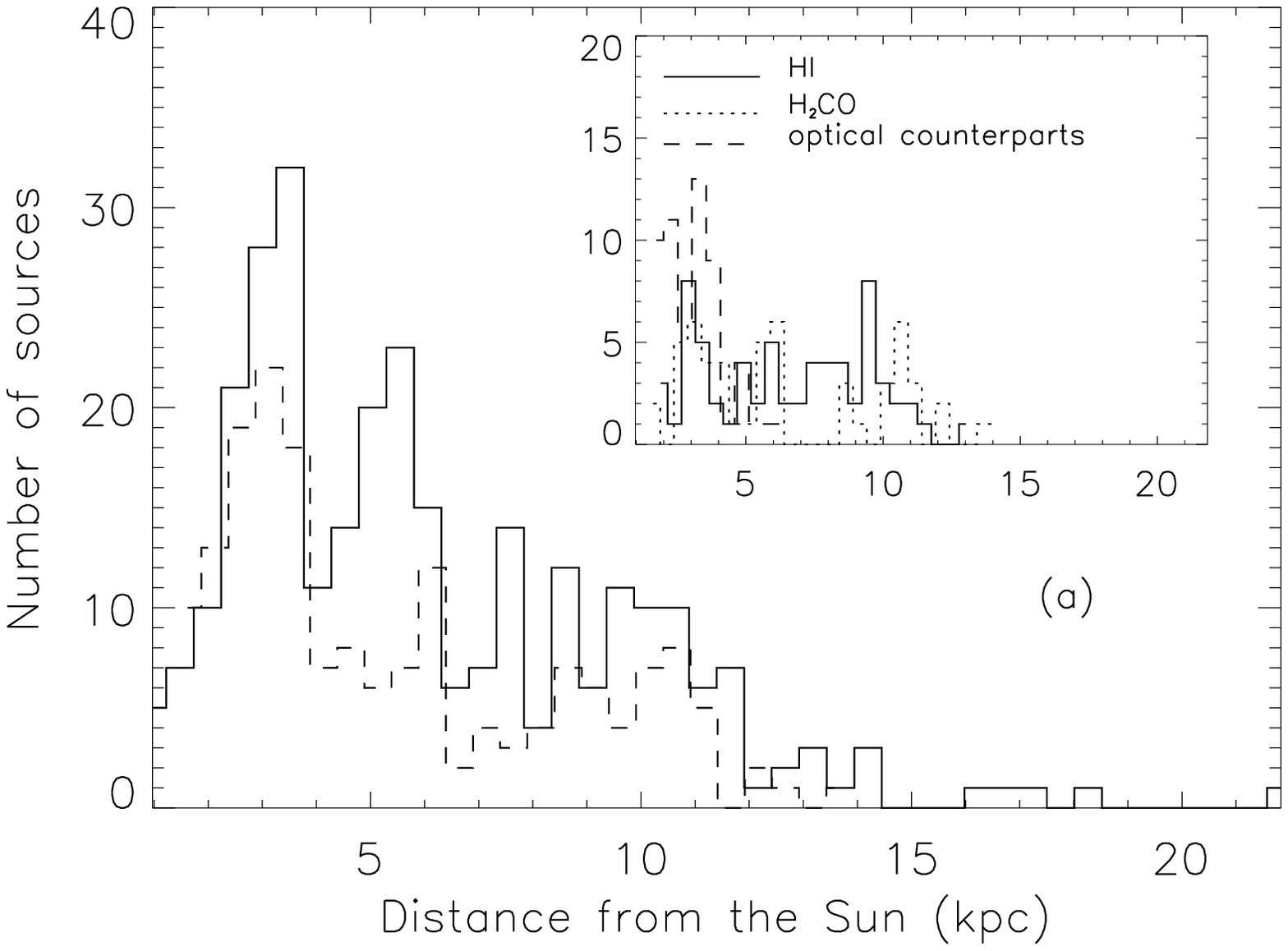,
height=5.5cm, width=8cm, angle=0}\\
\epsfig{figure=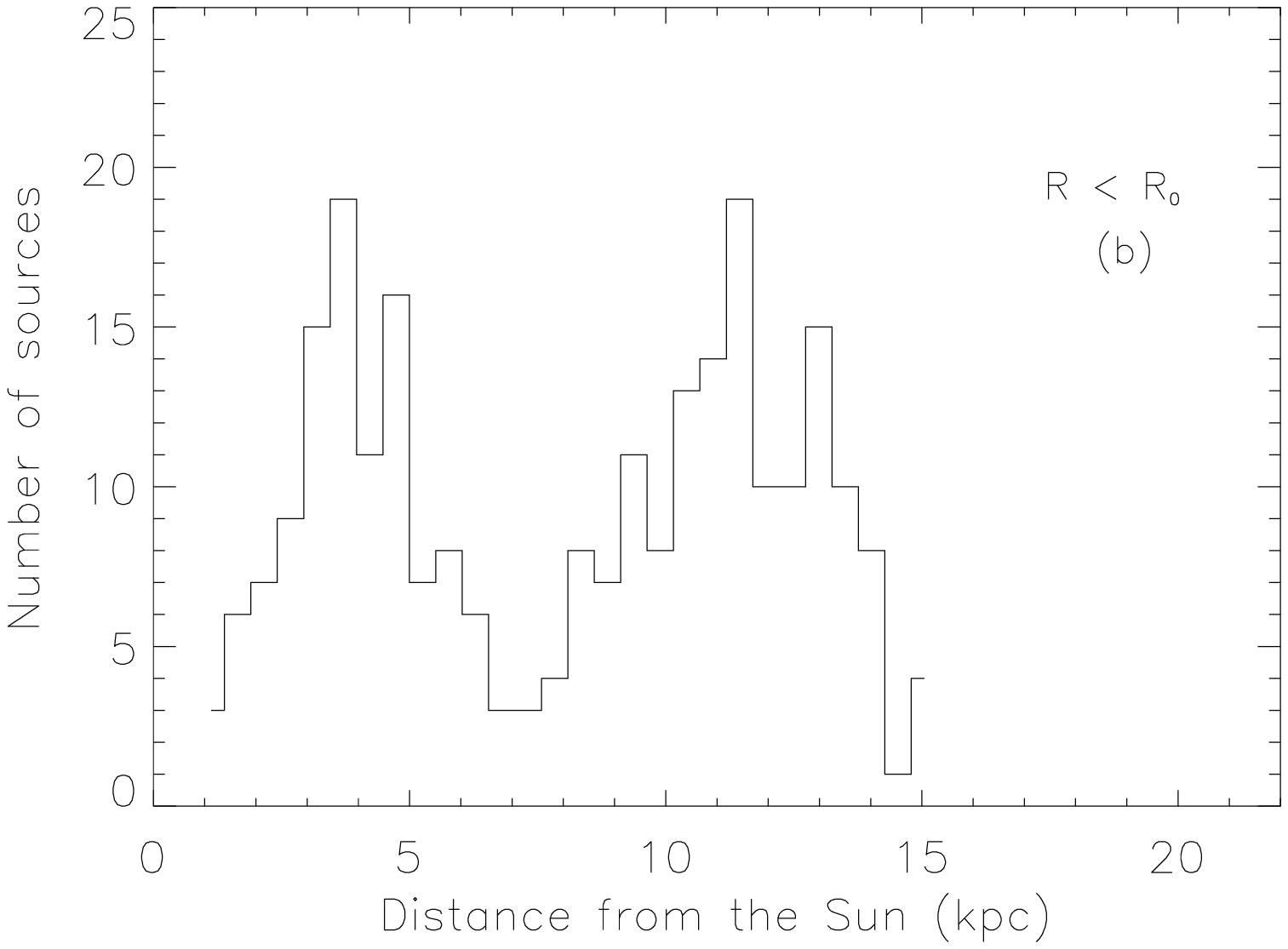,
height=5.5cm, width=8cm, angle=0}}
\caption{Distributions of solar
distances for HII regions. The top panel shows the distribution of
all 294 sources with unambiguous distance estimates (solid histogram)
and of the subset of 177 sources whose near/far degeneracy was broken
by complementary data (dashed). The insert details the distributions of
the various kinds of complementary data. Panel (b) shows the
distribution of 256 sources with near/far degeneracy resolved
using the luminosity-diameter relationship.}
\end{figure}

\begin{figure}
{\epsfig{figure=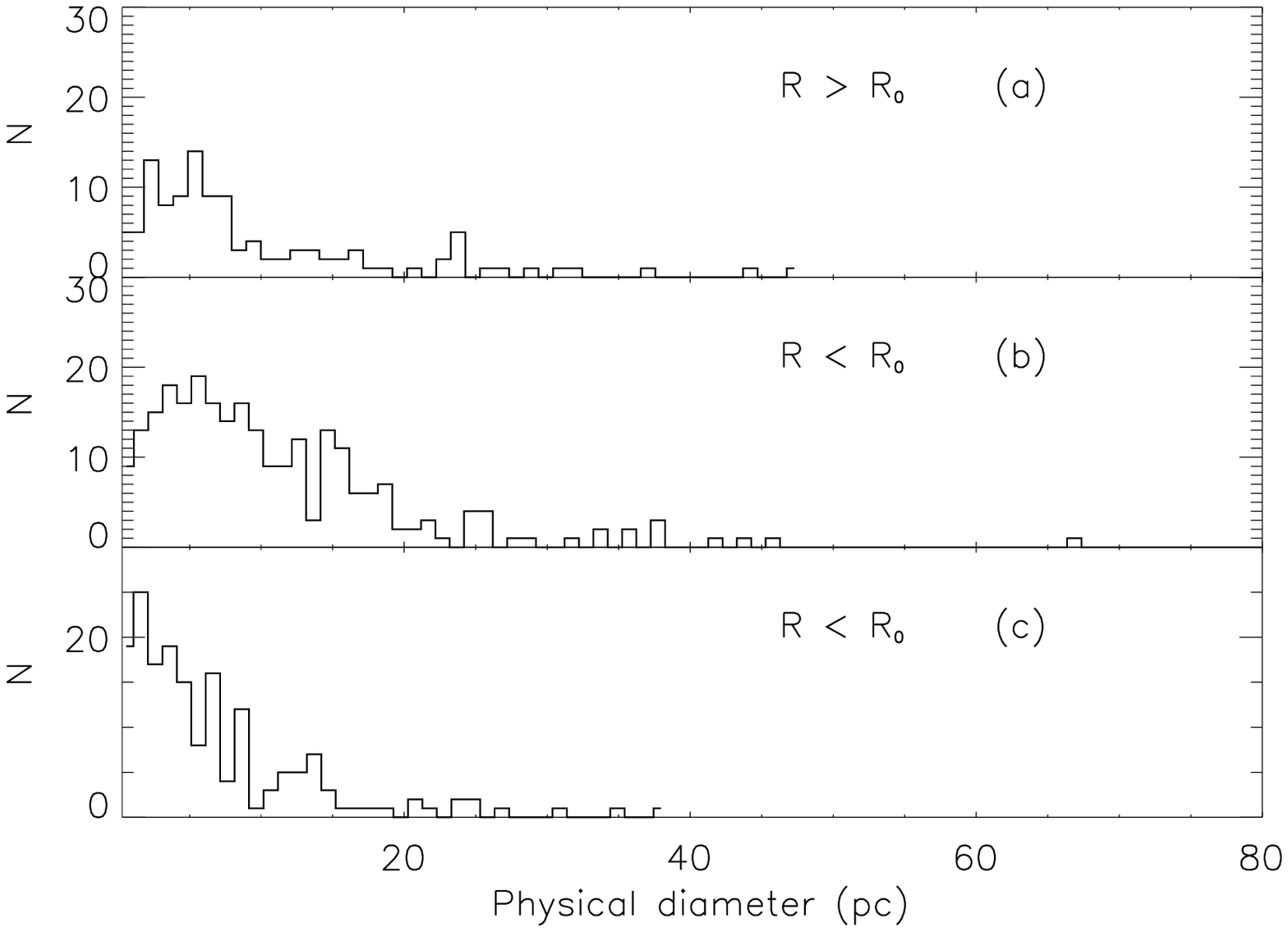,
height=8cm, width=8.cm, angle=0}} \caption{Distributions of linear
diameters. Panel (a): 117 sources with unique solutions for solar
distances; panel (b): 256 sources with near/far degeneracy resolved
using the luminosity-diameter relationship; panel (c): 177 sources
with near/far degeneracy resolved using auxiliary data.}
\end{figure}

\begin{figure}
{\epsfig{figure=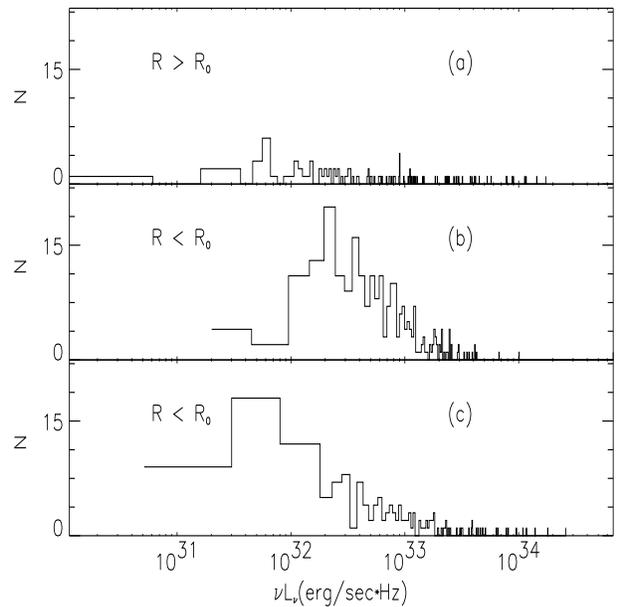,
height=8cm, width=8cm, angle=0}} \caption{Distributions of 2.7 GHz
luminosities. Panel (a): 117 sources with unique solutions for solar
distances; panel (b): 256 sources with near/far degeneracy resolved
using the luminosity-diameter relationship; panel (c): 177 sources
with near/far degeneracy resolved using auxiliary data.}
\end{figure}

\subsection[]{Distance degeneracy and properties of the sample}

\noindent
We will now consider the implications of breaking the distance
ambiguity on the properties of the HII regions lying within the solar circle
($R < R_{0}$). The considerations will include the distributions
of radial distance from the Sun, the physical diameter and the
source luminosity.

The distribution of distances from the Sun from different samples
with unambiguous distances are shown in Fig.~6(a). The dashed line is for the 177 sources with
$R < R_{0}$ whose distance ambiguity has been resolved using the
complementary data plotted in the insert. The full line is the sum of these
177 sources plus the 117 sources at $R > R_{0}$ with
unambiguous kinematic distances.

Fig.~6(b) shows the distributions of sources whose near/far
degeneracy has been resolved using the luminosity-diameter
relationship. It can be seen that the number of sources reaches a
first peak at 3-5 kpc from the Sun followed by a minimum at 6-9
kpc and a second peak at 10-14 kpc. The minimum corresponds to
distances from the Sun which include the Galactic centre region
where there is a deficit of detected sources.

The distribution of sources in Fig.~6(a) with $R < R_{0}$ (dashed
line) shows a strong deficit in the range 12-14 kpc compared with
Fig.~6(b). There are a number of contributing factors. Those
sources with optical counterparts are preferentially nearby
because of obscuration at larger distances. In addition, it is
more difficult to measure accurate absorption spectra for sources
at lower flux densities; these will again be preferentially at
larger distances. Table~3 illustrates this si\-tua\-tion. The
deeper survey by Kuchar $\&$ Bania (1994) shows a near to far
number ratio 20:72; 
while the less
deep survey by Caswell et al. (1975), Kerr $\&$ Knapp (1970) and
Wilson (1980) have a majority of sources at the near distance.

\begin{figure}
{\epsfig{figure=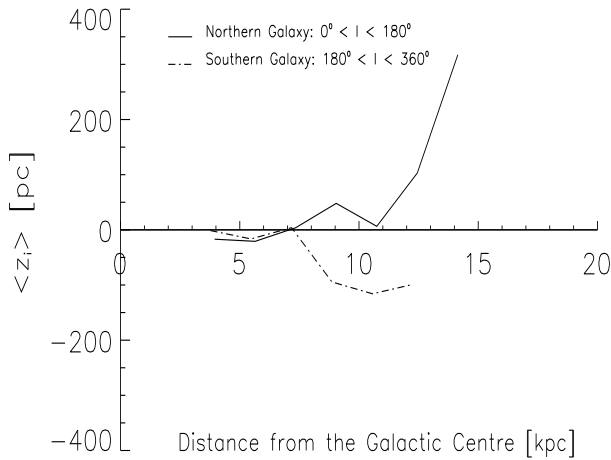,
height=6cm, width=8cm, angle=0}}
\caption{Variation of bin-averaged $z$ with distance from the Galactic
centre, $R$.}
\end{figure}

We now turn to the distribution of linear diameters deduced from
the application of our technique of resolving the distance
ambiguity. Low diameter sources are somewhat under-represented in
Fig.~7(b), since objects in this sample are on average at greater
distances as seen in Fig.~6(b). 
Fig.~8 shows the 2.7 GHz luminosity distribution for
the same three samples of sources as in Fig.~7. The 256 sources
(Fig.~8(b)) whose distance ambiguity is broken by the use of
Eq.~(7) have higher intrinsic luminosities than the other two
samples. Again this is because of their greater average distances 
for similar flux density and angular size distributions.

\subsection{The $z$-distribution of Galactic HII regions}

\noindent We take all 550 HII regions with distance determination
and estimate their $z$ distances from the Galactic plane using the
expression:
\begin{equation}
z = (D \times sin(b)) \times 10^{3}
\end{equation}
\noindent where $z$ is in pc, and the distance $D$ from the Sun is
in kpc. For the Northern (quadrants I and II; 0$^{0} \le l \le$
180$^{0}$) and the Southern (quadrants III and IV; 180$^{0} \le l
\le$ 360$^{0}$) Galaxy, we have binned the galactocentric
distances, $R$, into annuli 0.2$R_{0}$ wide. Fig.~9 shows the
mean of the $z$-distribution, $<z_{i}>$, where $i$ is the index of
the bin. The increasing $<z_{i}>$ with $R$ in quadrants I and II
and a corresponding decrease in quadrants III and IV clearly show
the warp of the Galactic plane well known from HI and CO studies,
although of smaller magnitude (see, for example, Burton 1988).

In order to investigate the thickening of the HII region layer
with $R$, we have also computed, for each bin, the width $\sigma_{i}$
of the $z$-distribution:
\begin{equation}
\sigma_{i} = \left[\sum_{j=1}^{N}(z_{j}-<z_{i}>)^{2}\right]^{\frac{1}{2}}
\end{equation}
where $N$ is the number of sources in the $i$-bin (see Table~4).
The uncertainty on $\sigma$ is dominated by the sampling error,
that we have estimated following Danese et al. (1980), on the
assumption of an underlying Gaussian distribution. The result is
illustrated in Fig.~10 for all 550 sources with distances and
shows that on both sides of the Galaxy, the HII region layer
thickens with increasing $R$. A similar effect is found for the
distribution of OB stars by Bronfman et al. (2000) and for the
distribution of molecular gas, based on the data from Cohen et al.
(1986), Grabelsky et al. (1987), May et al. (1988) and Digel
(1991), as reported in Bronfman et al. (2000).

\begin{figure}
{\epsfig{figure=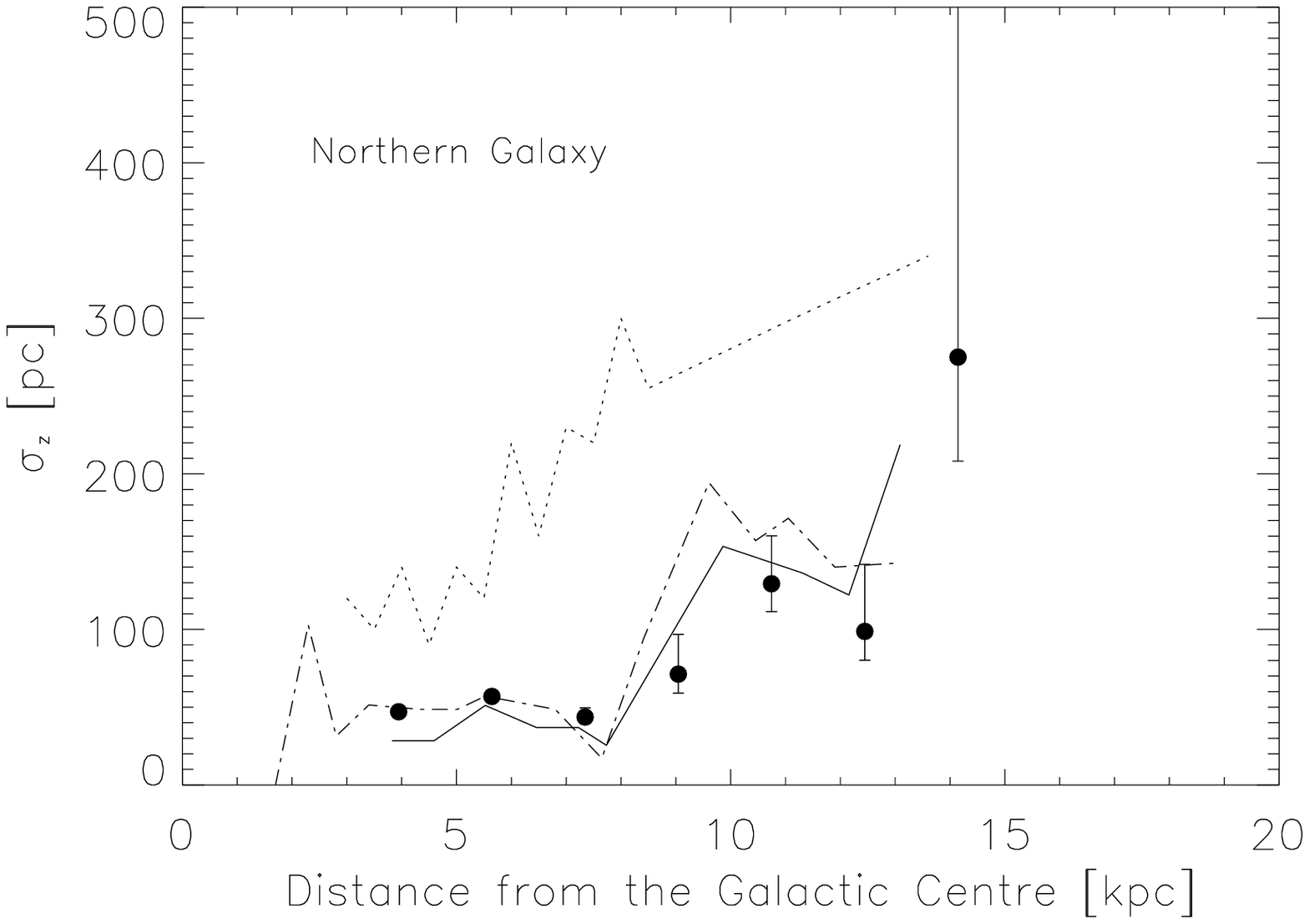,
height=6cm,width=8cm, angle=0}\\
\epsfig{figure=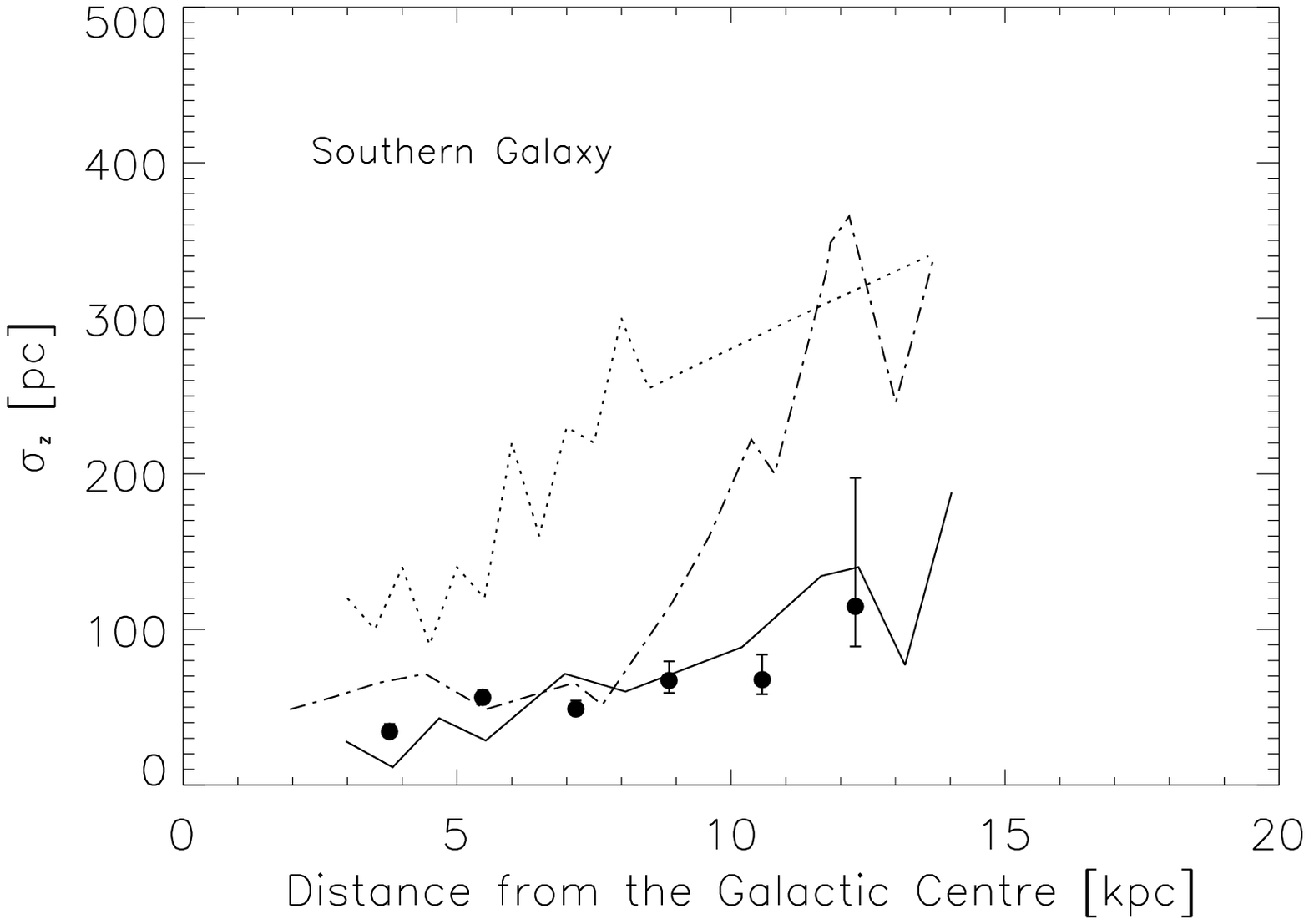,
height=6cm,width=8cm, angle=0}} \caption{Azimuthally-averaged
thickness in the Northern (0$^{0} \le l \le$ 180$^{0}$) and
Southern (180$^{0} \le l \le$ 360$^{0}$) Galaxy (filled circles).
Also shown for comparison are the UCHII layer (solid line), the
H$_{2}$ layer (dashed line) (both from Bronfman et al. 2000) and
the HI layer (dotted line). Data are from Malhotra (1995) for $R <
R_{0}$ and from Binney $\&$ Merrifield (1998) for $R > R_{0}$.}
\end{figure}

\begin{table}
\caption {Azymuthally-averaged $\sigma_{z}$ for the Northern and
the Southern hemisphere with sampling errors, under the assumption
of an underlying Gaussian distribution. Centroids of the bins ($R_{bin}$) and 
number of sources per bin ($N$) are also shown.}
\begin{center}
\begin{tabular}{c|cccc}\hline \hline
\multicolumn{5}{l}{\hskip 2.2truecm Northern \hskip 1.2truecm Southern}\\
\hline
    $R_{bin}$ (kpc)    &   N  &  $\sigma_{z}$ (pc)         &      N         &    $\sigma_{z}$ (pc)  \\
\hline
     3.9             &  103    &  47$^{+3.67}_{-2.97}$    &        36          &     34.3$^{+4.9}_{-3.5}$ \\
     5.6             & 124    &  56.9$^{+4}_{3.3}$        &       96          &    56.2$^{+4.6}_{-3.7}$              \\
     7.3             &  41     &  43.6$^{+5.9}_{4.2}$     &       57           &     48.8$^{+5.3}_{-4.0}$             \\
     9.0             &  10    &  71.2$^{+25.5}_{-12.2}$   &      24            &   67.1$^{+12.3}_{-7.9}$               \\
    10.7             &  19    & 129.3$^{+30.8}_{-17.9}$   &    15              &  67.6$^{+16.1}_{-9.4}$                \\
    12.4             &    8   &  98.7$^{+43}_{-18.5}$     &         5         &     114.7$^{+82.5}_{-25.7}$             \\
    14.1             &    4  & 275$^{+270}_{-66}$         &      -            &         -         \\
\hline\hline
\end{tabular}
\end{center}
\end{table}

For $R < R_{0}$, the width of the $z$-distribution (Fig.~10) is
almost constant. Table~5 gives the azimuthally averaged values of
$\sigma$ for the combined Northern and Southern data sets. The
best-determined distances give $\sigma \simeq 39.3\,$pc; when the
data for HII regions with distances relying on Eq.~(7) are
included, $\sigma \simeq 52\,$pc. These values are similar to the
values $\sigma = 32\,$pc derived for dust-embedded OB stars
(Bronfman et al. 2000) and $\sigma = 51\,$pc for the H$_{2}$ layer
(Bronfman et al. 1988). The HI (Malhotra 1995; Binney \&
Merrifield 1998) layer is wider by a factor of about 2. 
Pulsar dispersion measurements indicate a thin disk of ionized 
hydrogen with  $\sigma = 70$ pc (Reynolds 1991).  

\begin{table}
\caption {Width $\sigma$ and mean, $<z>$, of the $z$-distribution
within the solar circle. The results in the first column use
sources with unique solutions (u) plus sources whose distance
degeneracy was resolved with auxiliary data (aux) or with the
luminosity-diameter correlation (assig). In the second column, the
last group (assig) of sources is not included. }
\begin{center}
\begin{tabular}{ccccc}
\hline \hline
    $\sigma$ (pc)   &  $\sigma$ (pc)  &   $<z>$ (pc) \\
   u$+$aux$+$assig  &  u$+$aux       &                     \\
\hline
      52            &     -           &      -11.3             \\
       -            &     39.3          &     -7.3            \\
\hline\hline
\end{tabular}
\end{center}
\end{table}

The negative mean value of $z$, $<z>$, is consistent with the well
known result that the sun lies above the plane: Reed (1997) finds
$z_{\odot} = 10$--12 pc by analyzing the distribution of 12,522 OB
stars.

\begin{figure}
\centerline{\psfig{figure=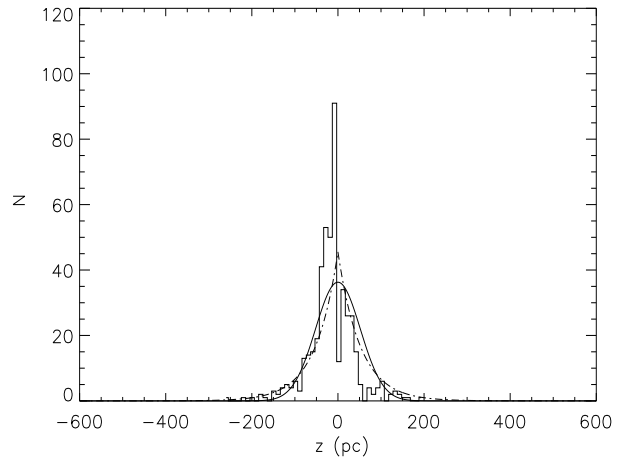,
height=6cm,width=8cm, angle=0}} \caption{z-distribution of the 456
sources at $R < R_{0}$ whose solar distance is either unambiguous
or assigned through the luminosity-diameter correlation. Overlaid
are a gaussian (solid line) and exponential curve (dashed line)
with $\sigma$ taken equal to 52 pc (see Table~5) and normalized to
the number of plotted sources. }
\end{figure}

\section{Conclusions}

\begin{figure*}
\epsfig{figure=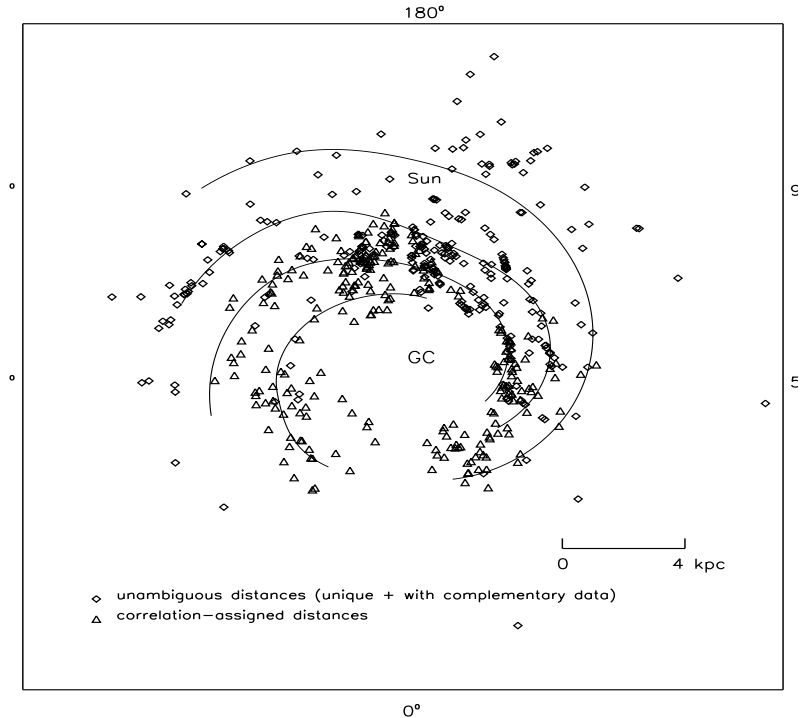,
height=9.5cm,width=10.5cm,angle=0} \caption{2-D distribution of
550 HII regions from the Catalog by Paladini et al. (2003). The
diamonds correspond to the 117 sources for which there is no
distance ambiguity, plus the 177 sources for which the distance
degeneracy was resolved thanks to auxiliary data. The triangles
correspond to the 256 sources to which the luminosity-diameter
correlation was applied. Also shown is the spiral arm model by
Taylor $\&$ Cordes (1993). An updated version of this model is
given in Cordes $\&$ Lazio 2002, 2003.}
\end{figure*}

\noindent We have analyzed the spatial distribution of 550
Galactic HII regions taken from the 2.7 GHz catalog of Paladini et
al. (2003), exploiting the extensive database of kinematic
information included in the catalog. For each source we have
derived a galactocentric distance using the rotation model by Fich
et al. (1989).

Distances from the Sun could be unambiguously derived for 117
sources, lying either outside the solar circle (84) or on a
line-of-sight tangential to their orbit (33). A highly significant
correlation between luminosity and linear diameter was found for
these sources. The corresponding least-square linear relationship
in the log-log plane was used to resolve, at least in a
statistical sense, the distance ambiguity for an additional 256
sources. The reliability of this approach was succesfully tested 
comparing the distributions of solar distances, linear diameters
and luminosities so obtained with those of sources
with unambiguous distances. 

An analysis of the $z$-distribution of HII regions shows:
\begin{enumerate}
\item{an increase of the mean value of $|z|$ with $R$, for $R>R_0$, reflecting the
shape of the warp;}
\item{a corresponding increase of the width of the distribution
as a function of $R$, comparable to what is seen for the OB stars
(Bronfman et al. 2000) and the molecular gas distribution (Cohen
et al. 1986; Grabelsky et al. 1987; May et al. 1988; Digel 1991);}
\item{an azimuthally-averaged thickness of the HII region
layer within the solar circle similar to that of OB stars
(Bronfman et al. 2000) but narrower than those of the diffuse HII
and HI (Reynolds 1991).}
\end{enumerate}

To check to what extent the above results depend on the adopted
linear diameter--luminosity correlation, we have repeated the
analysis adopting the extreme assumption that the two quantities
are totally independent. In this case we have used the mean linear
diameter ($\overline{d} = 7.6\,$pc) of sources with known solar
distance as a distance indicator. The results for $R > R_0$ (and,
in particular, the increase of $|<z>|$ and the thickening of the
$z$-width with increasing $R$) are obviously unchanged since
objects outside the solar circle are unaffected by the distance
degeneracy. For $R<R_0$ we find $\sigma = 51.4\,$pc, $<z> =
-7.2\,$pc, not significantly different from the values found using
the correlation (see Table~5).

Also, we confirm, for a much larger sample, the po\-si\-ti\-ve gradient
of electron temperature with galactocentric distance discovered in
previously published works.

The 2-D distribution within the Galactic plane of HII regions
having distances determined in the present study is shown in
Fig.~12. The HII regions show spiral-like
structures, in acceptable agreement with the spiral arms
delineated by Taylor $\&$ Cordes (1993, hereafter TC93). This fact
is partly expected due to the fact that the skeleton model of the
arm shapes in TC93 is derived from the map published in Georgelin
$\&$ Georgelin (1976), built on the basis of velocity data from
Reifenstein et al. (1970) and Wilson et al. (1970)) complemented by
the information contained in Downes et al. (1980) and Caswell $\&$
Haynes (1987). However, our analysis exploits data not available
for the TC93 analysis (see Table~1) and makes use of a different
rotation curve. From this point of view, the agreement with the
TC93 model is not a trivial result. A more detailed study of the
spiral arm structure is beyond the goal of the current work and it
will be presented in a forthcoming paper.

\section*{Acknowledgments}

R. Paladini thanks all the staff at the Jodrell Bank Observatory
for their hospitality and acknowledges financial support from a
Marie Curie Training Site Fellowhip. We are grateful to the
referee for constructive comments. Work supported in part by ASI
and MIUR.

\end{document}